\documentclass{article}
\usepackage{spconf,amsmath,graphicx}
\usepackage{multirow}
\usepackage{makecell}
\usepackage{caption} 
\usepackage{subcaption}
\usepackage{url}
\usepackage{tipa}
\title{Speech Recognition-based Feature Extraction for Enhanced Automatic Severity Classification in Dysarthric Speech}
\name{Yerin Choi$^*$, Jeehyun Lee$^*$, Myoung-Wan Koo$^\dagger$}	
\address{Department of Artificial Intelligence, Sogang University, Republic of Korea \\ 
\small{\texttt{\{lakahaga, jhlee22, mwkoo\}@sogang.ac.kr}}}
\begin{document}
%\ninept
%
\maketitle

\def\thefootnote{*}\footnotetext{Equal Contribution. Performed the work while at Sogang University.}
\def\thefootnote{†}\footnotetext{Corresponding Author.}
\def\thefootnote{\arabic{footnote}}

\begin{abstract}
Due to the subjective nature of current clinical evaluation, the need for automatic severity evaluation in dysarthric speech has emerged. DNN models outperform ML models but lack user-friendly explainability. ML models offer explainable results at a feature level, but their performance is comparatively lower. Current ML models extract various features from raw waveforms to predict severity. However, existing methods do not encompass all dysarthric features used in clinical evaluation. To address this gap, we propose a feature extraction method that minimizes information loss. We introduce an ASR transcription as a novel feature extraction source. We finetune the ASR model for dysarthric speech, then use this model to transcribe dysarthric speech and extract word segment boundary information. It enables capturing finer pronunciation and broader prosodic features. These features demonstrated an improved severity prediction performance to existing features: balanced accuracy of 83.72\%. 
\end{abstract}

\section{Introduction}
\label{sec:intro}

Dysarthria, a speech-motor disorder leading to significant speech intelligibility loss \cite{ImprovingKain}, can result from factors like stroke, tumors, Parkinson's Disease, and cerebral palsy. It impacts physical and psychological well-being, reducing the overall quality of life \cite{PatientsexperiencesDickson}. The primary cause of dysarthria is stroke, affecting around 50\% of stroke patients. 
Post-stroke dysarthria displays varied manifestations based on brain lesion location and size, highlighting the need for precise evaluation and personalized speech therapy \cite{mitchell2018feasibility}.
Current clinical practice heavily relies on resource-intensive perceptual auditory assessments by healthcare experts \cite{avan2019socioeconomic}. Moreover, the subjective nature of such evaluations raises reliability concerns. Perceptual severity assessment relies on analyzing characteristics in a patient's speech \cite{Wertz1992-pl}, necessitating continuous training for the involved experts.

Speech intelligibility is affected not only by the speaker's communication but also by the listener's comprehension \cite{Hustad2003-zf}. Consequently, there can be discrepancies in the evaluation of the same patient depending on the evaluator \cite{ListenerKate}. Specifically, experts such as doctors and speech pathologists, having extensive exposure to dysarthric speech, tend to rate speech intelligibility higher compared to the non-expert group \cite{dagenais2012acceptability}. Due to these limitations in perceptual evaluation \cite{MichaelaPerceptual}, the necessity for automated assessment has been emphasized \cite{zeplin1996reliability}.

With the rise of DNN, CNN and several speech self-supervised models automatically classified severity and showed performance improvements \cite{DysarthricZhengjun, AutomaticYeo, janbakhshi22_interspeech}. However, these models still lack the ability for real-world application due to their limited explainability. DNN models can only provide spectrum-based explanations (e.g., Grad-CAM \cite{gradcam}), which is hard for patients and healthcare experts to understand. DNN models struggle to interpret aspects influencing severity based on the perceptual assessment criteria commonly used in clinical practice. Another approach to automatic severity classification is ML classifiers with pre-defined features \cite{np18_interspeech, ClassificationQatab}. These ML-based classifiers can provide feature-level explanations about the result, which is more understandable for users. However, their performance is lower than that of DNN models when accurate prediction is key in the medical domain. 
In this paper, we propose a feature extraction method to enhance the performance of ML classifiers. Our method leverages the ML classifier's explainability and still obtains performance comparable to that of DNN models. Note that we do not propose a new ML classifier. Our focus is on the feature extraction method, applying features used in perceptual evaluations.

Previous ML-classifiers exploit information such as voice quality, prosody, acoustic, glottal, and phonetic features \cite{np18_interspeech,PhonologicalRudzicz,OROZCOARROYAVE2018207,TorgoSchu,xue21c_interspeech}. These features are obtained through signal-based calculation, often with toolkits like Praat \cite{boersma2001praat}. These cover various aspects of acoustic features of dysarthric speech. However, there are some unused aspects of speech in previous ML-classifiers that are used in clinical evaluation \cite{Duffy_2020}. The conventional feature extraction directly uses raw waveform as a source, resulting in a lack of semantic aspects.
For example, previous features only consider pause duration and the number of pause intervals, but in clinical evaluation, they also consider its locations.
This information loss in the feature extraction prevents us from incorporating all the features defined in terms of the medical perspective into the automatic diagnosis. 

We propose to employ the ASR (Automatic Speech Recognition) model as a source of features. We utilize a dysarthric ASR model to transcribe dysarthric speech and extract word segment boundaries. This allows us to capture fine-grained pronunciation correctness and utilize structural prosodic features. Plus, we design our speech recognition-based features according to the medically defined ones. Duffy \cite{Duffy_2020} categorized dysarthric speech features into five subsystems (respiration, phonation, articulation, resonance, and prosody), outlining specific attributes for each. Inspired by these attributes, we designed two main categories of features: \textit{Pronunciation Correctness}, and \textit{Structural Prosody}. 
\textit{Pronunciation Correctness} examine syntactic and semantic pronunciation and disfluency in the utterance using ASR transcription. 
Instead of formants, we evaluate pronunciation with transcription, enabling a detailed comparison with a reference.
\textit{Structural Prosody} investigates pause, articulation duration, and rhythm. It utilizes word segments obtained by ASR-based word segmentation. We construct diverse prosodic features employed by speech-language pathologists.
% Using ASR transcription and word segment boundaries, we extensively construct features for ML-classifier, obtaining 35 features in total. 

We applied the proposed method to a Korean dysarthric speech dataset. It contains paragraph reading speeches of post-stroke dysarthria patients. 
% DNN과의 비교
We compare the proposed method to both DNN and ML models. 
% integrated 
Additionally, we experimented with various combinations of previous and proposed features to analyze the efficacy of our feature extraction method. 
Multifaceted experiments show that our methods perform better than DNN models and previous feature extraction methods. 
Our speech recognition-based extraction method improves ML-classifier's performance, making it useful for explainable automatic dysarthria severity classification. We make the code for the proposed method public.\footnote{\url{https://github.com/lakahaga/ASR-dysarthria-classification}}

\section{Dataset}
\label{sec:dataset}

\begin{table}[h]
\centering
\caption{Dataset distribution according to severity scale.}
\label{tab:corpus-stats}
\resizebox{\columnwidth}{!}{%
\begin{tabular}{l|c|c|c|l}
\hline
\small{Severity} & \small{0} & \small{1} & \small{2} & \small{Total} \\ \hline \hline
\small{Number of Utterances} & \small{431} & \small{1950} & \small{186} & \small{2567} \\ \hline
\end{tabular}%
}
\end{table}

We employ the proposed method to the Korean Dysarthric Speech dataset\footnote{This paper used datasets from `The Open AI Dataset Project (AI-Hub, South Korea).' All data information can be accessed through `AI-Hub (www.aihub.or.kr)'}. 
It consists of recorded speech for several tasks used in clinical settings to evaluate the severity of dysarthria. Patients read the standard Korean paragraph for severity evaluation, referred to as the \textit{Autumn paragraph} \cite{kim2005dysarthria}. The \textit{Autumn paragraph} comprises six sentences designed to include the necessary consonants and vowels for the evaluation.
Trained healthcare professionals, such as neurosurgeons, annotated the severity levels. The severity scale follows the NIH Stroke Scale (NIHSS). NIHSS was used because this dataset collected post-stroke dysarthric speech.
The distribution of the dataset is shown in Table \ref{tab:corpus-stats}. Severity level 0 indicates no dysarthria, while severity levels 1 and 2 represent patients with dysarthria, with level 2 being the most severe.

\vspace{-0.2cm}

\begin{table*}[t]
\centering
\caption{Features for dysarthric severity classification based on Pronunciation Correctness and Structural Prosody analysis.}
\label{tab:new-features}
\resizebox{\textwidth}{!}{%
\begin{tabular}{c|c|ll}
\hline
\textbf{Main Category}              & \textbf{Secondary Category} & \multicolumn{2}{l}{\textbf{Features}}                                                                          \\ \hline \hline
                                    & Syntactic                   & \multicolumn{2}{l}{Insertion, Deletion, Substitution, WER, MER, WIL, WIP, Hits}                                \\ \cline{2-4} 
Pronunciation Correctness           & Semantic                    & \multicolumn{2}{l}{BERT score}                                                                                 \\ \cline{2-4} 
                                    & Disfluency                  & \multicolumn{2}{l}{Max Repetition, Filler Words Similarity}                                                    \\ \hline
\multirow{4}{*}{Structural Prosody} & Pause Location              & \multicolumn{2}{l}{Pause Insertion, Deletion, Substitution, CER, MER, WIL, WIP, Hits, DTW, Num}                           \\ \cline{2-4} 
                                    & Pause Duration              & \multicolumn{2}{l}{Pause Sum, Mean, SD (Standard Deviation), Max, Min}                                   \\ \cline{2-4} 
                                    & Articulation Duration       & \multicolumn{2}{l}{WS DTW, WS Duration statistics, Speech to Pause ratio, Top-30\% Short / Long WS} \\ \cline{2-4} 
                                    & Rhythm                      & \multicolumn{2}{l}{Abnormal Speed, Speed Change Rate Mean, Syllables Per Second, Increasing Speed}             \\ \hline
\end{tabular}%
}

\end{table*}

\section{Speech Recognition-based Feature Extraction}

We categorize speech recognition-based features (SR-features) into \textit{Pronunciation Correctness} and \textit{Structural Prosody}. Table \ref{tab:new-features} describes features we extracted from ASR transcription. We use the OpenAI's Whisper \cite{radford2022robust}. To accurately diagnose severity, we need an ASR model that can transcribe the pronunciation errors as they appear in dysarthric speech. The original Whisper tends to overestimate the pronunciation errors in dysarthric speech \cite{ip-detection}. Therefore, we finetune Whisper with dysarthric speech to transcribe dysarthric speech precisely. We refer to the finetuned Whisper as DysarthricWhisper. Using DysarthricWhisper, we obtain ASR transcription and word segment boundaries. For word segmentation, we used \textit{whisper-timestamped}\footnote{\url{https://github.com/linto-ai/whisper-timestamped}}. It uses cross-attention weights to get the segmentation information from the ASR results. Based on dynamic time warping and heuristics, it aligns audio with the inferred transcription.
% 여기에 icassp 를 쓰기! 
When fine-tuning Whisper, we added a special token for pause detection, as described in \cite{ip-detection}. In this way, we effectively transcribe and extract word segment information from dysarthric speech.

\subsection{Pronunciation Correctness}
\label{sec:feature-defs}

\textit{Pronunciation Correctness} measures the pronunciation errors in dysarthric speech compared to the original reading paragraph, \textit{Autumn paragraph}. We calculate these features using ASR transcription as hypothesis and original sentences in \textit{Autumn paragraph} as reference. \textit{Pronunciation Correctness} is further categorized into three subtypes. 

First, for \textit{Syntactic} correctness, we calculate typical metrics used in ASR performance evaluation, as listed in Table \ref{tab:new-features}. Second, \textit{Semantic} correctness is evaluated with BERT score \cite{bert-score}. It measures cosine similarity between tokens in hypothesis and reference sentence using contextual embeddings from the BERT \cite{bert}. BERT score strongly aligns with human assessments of ASR error types in disordered speech \cite{bert-score-dysarthric}. We utilize the KLUE-BERT \cite{klue-bert}. Lastly, for \textit{Disfluency}, we calculate \textit{Max Repetition} and \textit{Filler Words Similarity}. \textit{Max Repetition} measures the highest frequency of character repetitions to capture dysarthric patients' tendency toward repetitive speech patterns. \textit{Filler Words Similarity} assesses the occurrence of speech fillers (e.g., \textipa{[2]}, \textipa{[Wm]}, \textipa{[W]}, and \textipa{[kW]}). It is quantified by calculating cosine similarity between the final hidden states derived from the fillers and the hypothesis sentences using the KLUE-BERT base model.

\subsection{Structural Prosody} 
% pause랑 WS 뽑는 방법 설명
We extract three types of features for \textit{Structural Prosody}: \textit{Pause}, \textit{Articulation Duration}, and \textit{Rhythm}. As we added a pause token to Whisper, we could obtain pause duration by predicting the timestamps of special tokens.

% 각 feature 설명
\textit{Pauses} should appear at the right moment with the right amount, and patients with dysarthric find it hard to do it. Therefore, we look into the location of pauses, as well as the duration of pauses. In the case of the location of pauses, we collaborated with speech-language pathologists to annotate pauses commonly occurring in typical sentences within the \textit{Autumn paragraph}. The annotated pause location serves as the reference in the feature derivation.
Regarding location, we leverage ASR evaluation criteria again. We change ASR transcription with pause tokens into a binary sequence where 0 is for a word, and 1 is for a pause token. We refer to this binary sequence as a pause sequence. We calculate WER, CER, etc., with pause sequences. Plus, we measure \textit{Pause DTW} and \textit{the number of pauses}. \textit{Pause DTW} measures Dynamic Time Warping (DTW) distance to quantify the similarity between pause locations in the reference and hypothesis sequences. We also count how many pauses appeared. In the case of duration, we measure various statistics of durations of pauses that occurred in the input speech. 

Next, we consider \textit{Articulation Duration}. For articulation, we gather the segment boundaries of words from the predicted timestamps by \textit{whisper-timestamped} except for that of pause tokens. In this way, we can obtain the duration of articulations. From the duration sequence, we calculate DTW distance, \textit{WS (Word Segment) DTW}. The reference duration sequence is the average duration sequence of 360 samples read by healthy individuals. Also, we measure statistics of durations within speech to capture variability in durations among segments. To deal with various lengths of input speech, we normalize sum, mean, and sd with the number of segments. Other than basic statistics, we also analyze articulation duration with \textit{Speech to Pause Ratio} and \textit{Top 30\% Short / Long WS}. The \textit{Speech to Pause Ratio} is the ratio between the total duration of speech segments and the total duration of pauses, providing insights into the balance between speech and pauses in the given data. To include whether there is an anomaly of duration in the speech, we calculate the average duration of the top 30\% of the shortest / longest duration of word segments within a speech.  

Lastly, we construct \textit{Rhythm} features. \textit{Abnormal Speed} tends to check dysarthria-related anomalies in speech pacing. We compute the difference between the total duration of spoken segments and the average duration of the same sentence spoken by healthy individuals. 
While \textit{Abnormal Speed} focuses on each segment, \textit{Speed Change Rate Mean} measures the overall fluctuation in speaking pace. We calculate the average rate of change in duration between consecutive word segments. 
\textit{Increasing Speed} computes the average rate of change in duration between consecutive word segments, same as \textit{Speed Change Rate Mean}, but disregards negative changes. If the \textit{Speed Change Rate Mean} results in negative, \textit{Increasing Speed} is zero. In this way, \textit{Increasing Speed} checks whether an utterance shows speech tempo escalation from the beginning to the end of an utterance. 
\textit{Syllables Per Second (SPS)} is the average number of syllables spoken per second, disregarding pauses. It quantifies speech tempo in syllables inspired by WPM \cite{wpm}.

\section{Experiments}

\subsection{Experimental Setup}
\label{sec:experiment}
We used the same corpus to train the ASR system and automatic severity classification model. We split the Korean dysarthric speech dataset (Section \ref{sec:dataset}) into the train, validation, and test sets in the ratio of 8:1:1, considering the ratio of each severity level. The divided sets were maintained during ASR system training and classification model training. That being said, the samples in test sets have not been seen in training for both DysarthricWhisper and ML-classifier. We used TabularPredictor from AutoGluon-Tabular \cite{agtabular} for an ML classifier. The same trainer was used for the baseline and the proposed features. The use of AutoML ensures uniform experimentation conditions. 
% feature extraction
For feature extraction, we utilized the \texttt{jiwer}\footnote{\url{https://github.com/jitsi/jiwer}}  and \texttt{dtw-python}\footnote{\url{https://github.com/DynamicTimeWarping/dtw-python}} .
% ASR system training 
We used OpenAI's \texttt{whisper-small} \cite{radford2022robust} for the ASR system. When fine-tuning Whisper, AdamW \cite{adamw} was employed to optimize the model, with a learning rate of 5e-4 and batch size of 8. We utilized 2 A100-80GB GPUs.

\begin{table}[]
\centering
\caption{Raw Waveform-based features \cite{yeo2022cross}}
\label{tab:prev-features}
\resizebox{\columnwidth}{!}{%
\begin{tabular}{c|ll}
\hline
Category              & \multicolumn{2}{l}{Features}                                     \\ \hline \hline
\multirow{2}{*}{Voice Quality}                           & \multicolumn{2}{l}{Jitter, JitterDDP, JitterPPQ, Shimmer,} \\
                      & \multicolumn{2}{l}{ShimmerAPQ, HNR, LogHNR}                      \\ \hline
Articulation          & \multicolumn{2}{l}{F1, F2}                                       \\ \hline
Acoustic Prosody      & \multicolumn{2}{l}{Pitch, Intensity + overall statistics}        \\ \hline
\multicolumn{1}{l|}{\multirow{2}{*}{Structural Prosody}} & \multicolumn{2}{l}{Speech duration, Total duration,}       \\
\multicolumn{1}{l|}{} & \multicolumn{2}{l}{Pause duration, \# of pauses}                 \\ \hline
Speech rate           & \multicolumn{2}{l}{Speaking rate, Articulation rate, Phon ratio} \\ \hline
Rhythm           & \multicolumn{2}{l}{\%V, $\Delta V$, $\Delta C$, VarcoV, VarcoC, rPVIs, nPVIs} \\ \hline
\end{tabular}%
}
\end{table}

% baseline
We compared the proposed feature extraction method with an ML classifier, as well as previous features and DNN models. Previous features are listed in Table \ref{tab:prev-features}. We selected features based on \cite{yeo2022cross}. We used Forced aligner\footnote{\url{https://tutorial.tyoon.net/}} trained on Korean speech with human transcription for extracting \textit{Rhythm} features, and Parselmouth \cite{parselmouth} for others. The same AutoML classifier was employed.
For DNN models, we used two models, Whisper and wav2vec2.0 \cite{baevski2020wav2vec}, with two classifiers, the linear layer, and ResNet \cite{rathod23_interspeech}. We trained DNN models with the same data split setting. The AdamW was utilized, and the learning rate was set to 1e-3, with a batch size of 8 for both models.

% metrics

We used accuracy and balanced accuracy to evaluate the severity classification model.  The dataset we used is highly imbalanced, where about 75\% of test sets are severity 1. Therefore, we used balanced accuracy, which considers the class imbalance. Equation \ref{eq:bal-acc} and \ref{eq:sens-spec} describe balanced accuracy.
\vspace{-2pt}
\begin{equation}\
\resizebox{0.56\linewidth}{!}{$
    \text{\scriptsize{Balanced Accuracy}} = \frac{\text{Sensitivity} + \text{Specificity}}{2}
$}
\label{eq:bal-acc}
\end{equation}
\vspace{-12.6pt}
\begin{equation}\
\resizebox{0.65\linewidth}{!}{$
    \text{Sensitivity} = \frac{ TP}{(TP + FN)}, \text{Specificity} = \frac{ TN}{(TN + FP)}
$}
\label{eq:sens-spec}
\end{equation}

\subsection{Results}

Table \ref{tab:result} shows that the proposed feature extraction methodology (\textit{SR-features}) results in higher balanced accuracy compared to other feature extraction methods and deep learning models. In terms of accuracy, \textit{Whisper+Linear} showed the highest performance. However, as shown in Table \ref{tab:confusion-matrix}, \textit{Whisper+Linear} scored accuracy for severity 2 the lowest. Hence, balanced accuracy was lower compared to accuracy. As a matter of fact, balanced accuracy was higher in ML classifiers than in DNN models. DNN models were biased to severity 1, which was a major class in the dataset. In the case of pathologic speech, data scarcity, and class imbalance are prevalent. Although the automatic assessment model should be able to handle these challenges effectively, the DNN models lack those kinds of capabilities. \textit{Waveform} feature set also showed higher balanced accuracy than DNN models, but the accuracy for severity 2 was still 44.44\%. Conversely, our method improved performance in both minor classes, such as severity 0 and 2, especially severity 2. Plus, the case of Rhythm in Table \ref{tab:prev-features} requires human transcription, whereas the proposed features use ASR transcription.
Even though \textit{SR-features} only covers pronunciation and structural prosody, and the \textit{Waveform} set contains acoustic features as well, \textit{SR-features} showed the highest balanced accuracy. 
% asr model이 쉰 소리가 나거나, voice break가 있으면 해당 부분에서의 transcrption에서도 오류가 나기 때문에? 이를 implicit하게 반영한다고 볼 수 있다? 

\begin{table}[t]
\centering
\caption{Automatic severity classification model performance.}
\label{tab:result}
\resizebox{\columnwidth}{!}{%
\renewcommand{\arraystretch}{1.3}
\begin{tabular}{c||l|l||l|l}
\hline 
\multirow{5}{*}{\textbf{DNN}} & \textbf{Model}                       & \textbf{Classifier} & \textbf{Accuracy} & \textbf{Balanced Accuracy} \\ \cline{2-5} \noalign{\vskip\doublerulesep
         \vskip-\arrayrulewidth} \cline{2-5}
                     & \multirow{2}{*}{Whisper}    & Linear     & \textbf{91.94}    & 61.28             \\ \cline{3-5} 
                     &                             & ResNet \cite{rathod23_interspeech}    & 69.89    & 53.69             \\ \cline{2-5} 
                     & \multirow{2}{*}{wav2vec2.0} & Linear     & 91.13    & 62.47          \\ \cline{3-5} 
                     &                             & ResNet     & 82.53    & 57.90          \\ \hline \hline 
\multirow{3}{*}{\textbf{ML}}  & \textbf{Feature set}                 & \textbf{Classifier} & \textbf{Accuracy} & \textbf{Balanced Accuracy} \\ \cline{2-5} \noalign{\vskip\doublerulesep
         \vskip-\arrayrulewidth} \cline{2-5}
                     & Waveform                    & \multirow{2}{*}{AutoML}    & 81.72    & 68.96             \\  \cline{2-2} \cline{4-5} 
                     & SR-features (Ours)   &     & 72.85    & \textbf{83.72}             \\ \hline 
\end{tabular}%
}
\end{table}

\begin{table}[t]
\centering
\caption{Confusion matrices of each model}
\label{tab:confusion-matrix}
 % \caption{Confusion matrices of each model}
     \begin{subtable}{\columnwidth}
        \centering
        \caption{Whisper + Linear}
        \resizebox{\columnwidth}{!}{%
        \begin{tabular}{l | l | l | l}
        \hline
        GT \textbackslash Pred & 0 & 1 & 2 \\
        \hline \hline
        0 & \textbf{85.19\% (46/54)} & 14.81\% (8/54) & 0.00\% (0/54)\\ \hline
        1 & 1.33\%(4/300) & \textbf{98.67\% (296/300)} & 0.00\%\ (0/300)\\ \hline
        2 & 0.00\% (0/18) & 100.00\% (18/18) & \textbf{00.00\% (0/18)}\\ \hline
        \end{tabular}%
        }
    \vspace{2mm}
    \end{subtable}
    \vspace{2mm}
    \begin{subtable}{\columnwidth}
        \centering
        \caption{Waveform Features}
        \resizebox{\columnwidth}{!}{%
        \begin{tabular}{l | l | l | l}
        \hline
        GT  \textbackslash Pred & 0 & 1 & 2 \\
        \hline \hline
        0 & \textbf{77.78\% (42/54)} & 22.22\% (12/54) & 0.00\% (0/54)\\ \hline
        1 & 10.33\%(31/300) & \textbf{84.67\% (254/300)} & 5.00\%\ (15/300)\\ \hline
        2 & 0.00\% (0/18) & 55.56\% (10/18) & \textbf{44.44\% (8/18)}\\ \hline
       \end{tabular}%
        }
    \end{subtable}

    \begin{subtable}{\columnwidth}
        \centering
        \caption{Ours: SR-features}
        \resizebox{\columnwidth}{!}{%
        \begin{tabular}{l | l | l | l}
        \hline
        GT \textbackslash Pred & 0 & 1 & 2 \\
        \hline \hline
        0 & \textbf{81.48\% (44/54)} & 18.52\% (10/54 ) & 0.00\% (0/54)\\ \hline
        1 & 17.67\%(53/300) & \textbf{69.67\% (209/300)} & 12.67\%\ (38/300)\\ \hline
        2 & 0.00\% (0/18) & 0.00\% (0/18) & \textbf{100.00\% (18/18)}\\ \hline
       \end{tabular}%
        }
     \end{subtable}
\end{table}

\begin{table}[t]
\caption{ML-Classifier performance of each category}
\label{tab:ablation}
\resizebox{\columnwidth}{!}{%
\begin{tabular}{l|l|l}
\hline
                     & Accuracy & Balanced Accuracy \\ \hline \hline
SR-features (Ours)  & 72.85    & \textbf{83.72}     \\ \hline \hline
Only Structural Prosody   & \textbf{74.19} & 81.80               \\ \hline
Only Pronunciation Correctness  & 57.53 & 78.40                    \\ \hline
\end{tabular}%
}
\end{table}

\begin{table}[t]
\centering
\caption{Confusion matrices of each category}
\label{tab:abl-confusion-matrix}
     \begin{subtable}{\columnwidth}
        \centering
        \caption{\textit{Structural Prosody}}
        \resizebox{\columnwidth}{!}{%
        \begin{tabular}{l | l | l | l}
        \hline
        GT \textbackslash Pred & 0 & 1 & 2 \\
        \hline \hline
        0 & \textbf{85.19\% (46/54)} & 14.81\% (8/54) & 0.00\% (0/54)\\ \hline
        1 & 21.33\%(64/300) & \textbf{71.33\% (214/300)} & 7.33\%\ (22/300)\\ \hline
        2 & 0.00\% (0/18) & 11.11\% (2/18) & \textbf{88.89\% (16/18)}\\ \hline
       \end{tabular}%
        }
    \vspace{2mm}
    \end{subtable}
    
    \begin{subtable}{\columnwidth}
        \centering
        \caption{\textit{Pronunciation Correctness}}
        \resizebox{\columnwidth}{!}{%
        \begin{tabular}{l | l | l | l}
        \hline
        GT  \textbackslash Pred & 0 & 1 & 2 \\
        \hline \hline
        0 & \textbf{85.19\% (46/54)} & 11.11\% (6/54) & 3.70\% (2/54)\\ \hline
        1 & 38.00\%(114/300) & \textbf{50.00\% (150/300)} & 12.00\%\ (36/300)\\ \hline
        2 & 0.00\% (0/18) & 0.00\% (0/18) & \textbf{100.00\% (18/18)}\\ \hline
       \end{tabular}%
        }
    
    \end{subtable}
    
\end{table}

\subsection{Ablation Study}

We dissect the effect of the proposed method by training ML-classifiers with each main category in Table \ref{tab:new-features}. We used the same AutoML classifier described in Section \ref{sec:experiment}. Table \ref{tab:ablation} shows the classification performance according to each feature category: \textit{Structural Prosody} and \textit{Pronunciation Correctness}. Plus, Table \ref{tab:abl-confusion-matrix} shows a detailed evaluation with confusion matrices.
The balanced accuracies of both subsets are higher than the baselines in Table \ref{tab:result}. \textit{Pronunciation correctness} showed better accuracy in severity 0 and 2 than other models while scoring only 50\% in severity 1. 
\textit{Structural Prosody} showed a marginal performance difference from the proposed feature set (\textit{SR-features}). Furthermore, we calculated the feature importance with the result of the proposed method (\textit{SR-features}). The top 5 ranked features belonged to \textit{Structural Prosody}. (Top 5 Features: \textit{Speed Change Rate Mean, Syllables Per Second, WS Duration Sum, Top-30\% Short WS, Increasing Speed}) We used permutation-based feature importance calculation.
% sturucutal prosody가 가장 높은 balanced accuracy를 보였고, pronunciation correctness는 sev0,2에서는 좋은 성능을 보였지만, sev1에서 성능 저하를 보였다. 또한 sev1에서는 Structural prosody의 성능이 더 좋았으며, 전체적인 성능도 전체 feature와 크게 차이 나지 않았다. 또한, test set에서 feature importance를 계산하였을 때 상위 5개의 feature가 모두 structural prosody에 해당했다. 
% Speech recognition based feature에서 성능 향상에 많은 영향을 끼친 것은 Structural prosoyd feature들임을 알 수 있었다. 

\subsection{Integration of feature sets}
In this additional experiment, we look into the impact of the number of features and feature coverage.
Since the \textit{SR-features} set has more features than the ML baseline, the improved performance could stem from the number of input features. 
Furthermore, \textit{SR-features} only over pronunciation and structural prosody, while \textit{Waveform} features cover broader aspects of speech. 
Therefore, we compare the proposed method with the integrated feature set. 
The integrated feature set contains all features in Table \ref{tab:new-features} and Table \ref{tab:prev-features}.
We compare three combinations of features with the proposed method. First, we simply combine two feature sets. (\textit{Integerated-without-FS} in Table \ref{tab:fs})
Then, we conducted auto feature selection\footnote{\url{https://github.com/AutoViML/featurewiz.git}}. We run feature selection for each feature set respectively and then integrate only the selected features, referred to as \textit{Integrated-after-FS} in Table \ref{tab:fs}. \textit{Integarted-before-FS} indicates the set integrated all features first, then ran auto feature selection. We used the same AutoML classifier. 
% The optimal feature set for each selection is listed in Table \ref{tab:optimal-features}. 

Employing many more features resulted in improved accuracy. However, the balanced accuracy was still lower than the proposed method. This means our method performs better, not because of the higher number of features. \textit{Integrated-wihtout-FS} had more features than \textit{SR-features} set but resulted in lower balanced accuracy.
Feature selection was also ineffective in terms of performance. This shows that ASR results are a resourceful source of features for automatically assessing the severity of dysarthria. 

\begin{table}[t]
\caption{ML-Classifier performance of \textit{Integrated} sets. \textit{W} is the number of Waveform features in \textit{Integrated} set, and \textit{S} is that of the \textit{SR-features} set.}
\label{tab:fs}
\resizebox{\columnwidth}{!}{%
\begin{tabular}{l|l|l|l}
\hline
              & \# of features  (W + S)        & Accuracy & Balanced Accuracy \\ \hline \hline
              
SR-features Only (Ours) &  42 & 72.85    & \textbf{83.72}     \\ \hline \hline
Integrated-without-FS  & 74 (32 + 42) & \textbf{88.98} & 76.46                  \\ \hline
Integrated-before-FS  &  38  (15 + 23) & 77.69 & 77.01                    \\ \hline
Integrated-after-FS  &   39  (17 + 22) &  83.06 & 77.49                  \\  \hline
\end{tabular}%
}
\end{table}
\vspace{-2pt}

\begin{figure}[h]
\centering
    \includegraphics[width=\columnwidth]{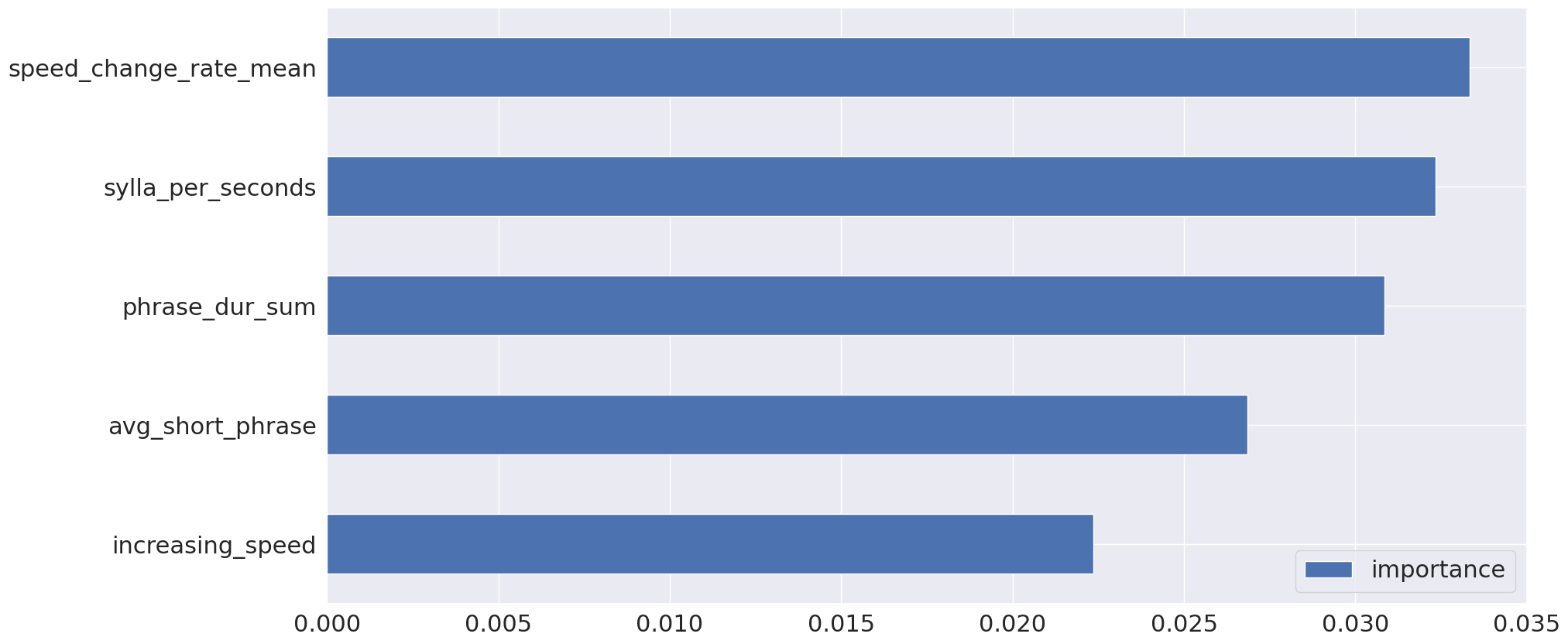}
    \caption{Top 5 most important features of the proposed model tested on the test set.}
\label{fig:feature-importance}
\end{figure}

\begin{figure}[th]
    \centering
    \begin{subfigure}[b]{0.42\textwidth}
        \centering
        \includegraphics[width=\columnwidth]{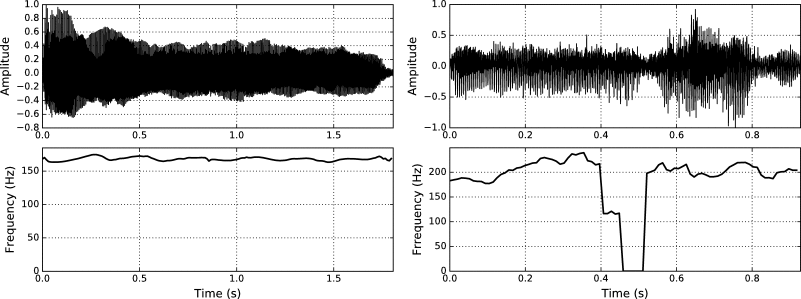}
        \caption{Example of result explanation for \textit{Waveform}}
        \vspace{2mm}
    \end{subfigure}
    \hfill
    \begin{subfigure}[b]{0.42\textwidth}
        \centering
        \includegraphics[width=\columnwidth]{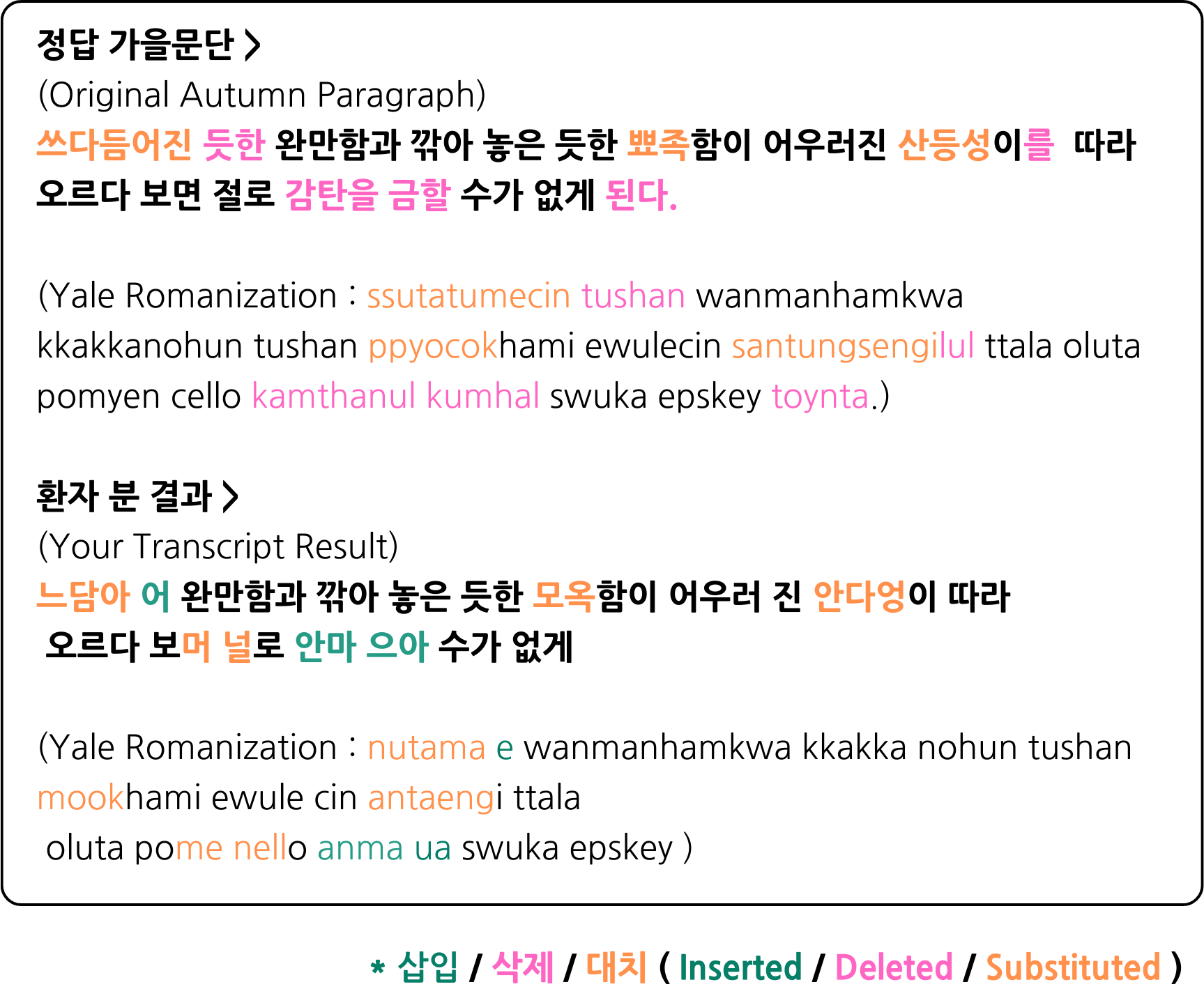}
        \caption{Example of result explanation for \textit{ASR Transcription}}
    \end{subfigure} 
    \caption{Examples of result explanation. The top-side illustration compares speech signals and fundamental frequencies between a healthy speaker and a Parkinson's Disease patient. On the bottom, an ASR-based illustration highlights character-level changes in speech. The upper part of the bottom-side illustration represents a healthy speaker's reading, while the lower part displays the ASR model's inferred outcome for a patient.}
    \label{fig:explainability}
\end{figure}

\section{Discussion}
% 속도
% sev2에 대한 성능은 높였지만, sev1에 대한 성능은 낮아짐. severity 별로 에측 성능이 달라지는 문제점이 있음.  하지만 이전 모델들에 비해 그 차이는 줄어듦. 이러한 문제점은 데이터셋의 불균형을 해소하는 방법?
Our approach can offer feature-level analysis of the predicted severity, providing insight into the patient's severity and speech therapy planning. For example, as in Figure \ref{fig:feature-importance}, our method could reveal the impact on severity through feature importance analysis aligned with clinical indicators. 
Moreover, our method can provide an easily comprehendible text-based explanation of the proposed ASR-based features. We demonstrated the potential for offering explanations through Figure \ref{fig:explainability}. The illustration on the top \cite{OROZCOARROYAVE2018207} presents a comparison of speech signals and fundamental frequencies during sustained vowel  `\textit{A}' phonation. The left side corresponds to a healthy speaker, while the right side represents a Parkinson's Disease patient. In contrast, bottom-side illustration is based on ASR Transcription, highlighting character-level insertions, deletions, and substitutions in speech utterances. This version offers a more accessible explanation for a general audience, aiding their comprehension of these alterations in dysarthric speech. It also emphasizes that patients, including those with Parkinson's Disease, can use this information for training and improvement. This visual guide helps individuals understand the specific aspects they should focus on while practicing speaking.

While the proposed method shows improved performance in minor classes, accuracy in severity 1 has degraded. Although improving performance in severity 2 has reduced the performance difference across classes, the disparity remains. This performance still differed across classes when we used class-imbalance solutions like weighted loss. We believe that improving dysarthric ASR will enhance the quality of the proposed features, which will alleviate this limitation. (CER of \textit{DysarthricWhisper}: 11.96\%)

\section{Conclusion}

We propose the speech recognition-based feature extraction method for dysarthric speech's automatic severity classification. We employed an ML model, which has feature-level interpretability and improved its performance. To this end, we quantified the clinically utilized dysarthric features. We fine-tuned the ASR model with dysarthric speech to effectively transcribe the dysarthric speech for extended feature extraction. With DysarthricWhisper, we transcribe extract word segment boundaries and detect pauses. We construct \textit{Pronunciation Correctness} and \textit{Structural Prosody} features from extended ASR results. We proved the enhanced performance compared to previous DNN and ML models. With the ablation study, we further analyzed the impact of each feature category. Even though the proposed feature set covers fewer aspects than the previous features and integrated features, it showed better balanced accuracy. As the major contribution, we constructed efficient features from ASR transcription and word segment information, especially improving its accuracy for severity 2. Our method highlights that a dysarthric speech fine-tuned ASR model is resourceful for severity prediction.

\section{Acknowledgment}

This work was supported by Institute of Information \& communications Technology Planning \& Evaluation (IITP) grant funded by the Korea government(MSIT) (No.2022-0-00621, Development of artificial intelligence technology that provides dialog-based multi-modal explainability)

% Below is an example of how to insert images. Delete the ``\vspace'' line,
% uncomment the preceding line ``\centerline...'' and replace ``imageX.ps''
% with a suitable PostScript file name.
% -------------------------------------------------------------------------
% \begin{figure}[htb]
	
% 	\begin{minipage}[b]{1.0\linewidth}
% 		\centering
% 		\centerline{\includegraphics[width=8.5cm]{image1}}
% 		%  \vspace{2.0cm}
% 		\centerline{(a) Result 1}\medskip
% 	\end{minipage}
% 	%
% 	\begin{minipage}[b]{.48\linewidth}
% 		\centering
% 		\centerline{\includegraphics[width=4.0cm]{image3}}
% 		%  \vspace{1.5cm}
% 		\centerline{(b) Results 3}\medskip
% 	\end{minipage}
% 	\hfill
% 	\begin{minipage}[b]{0.48\linewidth}
% 		\centering
% 		\centerline{\includegraphics[width=4.0cm]{image4}}
% 		%  \vspace{1.5cm}
% 		\centerline{(c) Result 4}\medskip
% 	\end{minipage}
% 	%
% 	\caption{Example of placing a figure with experimental results.}
% 	\label{fig:res}
% 	%
% \end{figure}

% References should be produced using the bibtex program from suitable
% BiBTeX files (here: strings, refs, manuals). The IEEEbib.bst bibliography
% style file from IEEE produces unsorted bibliography list.
% -------------------------------------------------------------------------
\bibliographystyle{IEEEbib}
\bibliography{main}

\begin{thebibliography}{10}

\bibitem{ImprovingKain}
Alexander~B. Kain, John-Paul Hosom, Xiaochuan Niu, Jan~P.H. {van Santen}, Melanie Fried-Oken, and Janice Staehely,
\newblock ``Improving the intelligibility of dysarthric speech,''
\newblock {\em Speech Communication}, vol. 49, no. 9, pp. 743--759, 2007.

\bibitem{PatientsexperiencesDickson}
Sylvia Dickson, Rosaline~S. Barbour, Marian Brady, Alexander~M. Clark, and Gillian Paton,
\newblock ``Patients' experiences of disruptions associated with post-stroke dysarthria,''
\newblock {\em International Journal of Language \& Communication Disorders}, vol. 43, no. 2, pp. 135--153, 2008.

\bibitem{mitchell2018feasibility}
Claire Mitchell, Audrey Bowen, Sarah Tyson, and Paul Conroy,
\newblock ``A feasibility randomized controlled trial of readyspeech for people with dysarthria after stroke,''
\newblock {\em Clinical Rehabilitation}, vol. 32, no. 8, pp. 1037--1046, 2018.

\bibitem{avan2019socioeconomic}
Abolfazl Avan, Hadi Digaleh, Mario Di~Napoli, Saverio Stranges, Reza Behrouz, Golnaz Shojaeianbabaei, Amin Amiri, Reza Tabrizi, Naghmeh Mokhber, J~David Spence, et~al.,
\newblock ``Socioeconomic status and stroke incidence, prevalence, mortality, and worldwide burden: an ecological analysis from the global burden of disease study 2017,''
\newblock {\em BMC medicine}, vol. 17, no. 1, pp. 1--30, 2019.

\bibitem{Wertz1992-pl}
Robert Wertz and John Rosenbek,
\newblock ``Where the ear fits: A perceptual evaluation of motor speech disorders,''
\newblock {\em Semin. Speech Lang.}, vol. 13, no. 01, pp. 39--54, Feb. 1992.

\bibitem{Hustad2003-zf}
Katherine~C Hustad, Tabitha Jones, and Suzanne Dailey,
\newblock ``Implementing speech supplementation strategies: effects on intelligibility and speech rate of individuals with chronic severe dysarthria,''
\newblock {\em J. Speech Lang. Hear. Res.}, vol. 46, no. 2, pp. 462--474, Apr. 2003.

\bibitem{ListenerKate}
Kate Bunton, Raymond~D. Kent, Joseph~R. Duffy, John~C. Rosenbek, and Jane~F. Kent,
\newblock ``Listener agreement for auditory-perceptual ratings of dysarthria,''
\newblock {\em Journal of Speech, Language, and Hearing Research}, vol. 50, no. 6, pp. 1481--1495, 2007.

\bibitem{dagenais2012acceptability}
Paul~A Dagenais and Amy~F Wilson,
\newblock ``Acceptability and intelligibility of moderately dysarthric speech by four types of listeners,''
\newblock in {\em Investigations in clinical phonetics and linguistics}, pp. 379--388. Psychology Press, 2012.

\bibitem{MichaelaPerceptual}
Michaela Pernon, Frédéric Assal, Ina Kodrasi, and Marina Laganaro,
\newblock ``Perceptual classification of motor speech disorders: The role of severity, speech task, and listener's expertise,''
\newblock {\em Journal of Speech, Language, and Hearing Research}, vol. 65, no. 8, pp. 2727--2747, 2022.

\bibitem{zeplin1996reliability}
Joslin Zeplin and Ray~D Kent,
\newblock ``Reliability of auditory-perceptual scaling of dysarthria,''
\newblock {\em Disorders of motor speech: Assessment, treatment, and clinical characterization}, pp. 145--154, 1996.

\bibitem{DysarthricZhengjun}
Zhengjun Yue, Erfan Loweimi, and Zoran Cvetkovic,
\newblock {\em Dysarthric Speech Recognition, Detection and Classification using Raw Phase and Magnitude Spectra},
\newblock ISCA-INST SPEECH COMMUNICATION ASSOC, May 2023.

\bibitem{AutomaticYeo}
Eun~Jung Yeo, Kwanghee Choi, Sunhee Kim, and Minhwa Chung,
\newblock ``Automatic severity classification of dysarthric speech by using self-supervised model with multi-task learning,''
\newblock in {\em ICASSP 2023 - 2023 IEEE International Conference on Acoustics, Speech and Signal Processing (ICASSP)}, 2023, pp. 1--5.

\bibitem{janbakhshi22_interspeech}
Parvaneh Janbakhshi and Ina Kodrasi,
\newblock ``{Adversarial-Free Speaker Identity-Invariant Representation Learning for Automatic Dysarthric Speech Classification},''
\newblock in {\em Proc. Interspeech 2022}, 2022, pp. 2138--2142.

\bibitem{gradcam}
Ramprasaath~R. Selvaraju, Michael Cogswell, Abhishek Das, Ramakrishna Vedantam, Devi Parikh, and Dhruv Batra,
\newblock ``Grad-cam: Visual explanations from deep networks via gradient-based localization,''
\newblock in {\em 2017 IEEE International Conference on Computer Vision (ICCV)}, 2017, pp. 618--626.

\bibitem{np18_interspeech}
Narendra {N P} and Paavo Alku,
\newblock ``{Dysarthric Speech Classification Using Glottal Features Computed from Non-words, Words and Sentences},''
\newblock in {\em Proc. Interspeech 2018}, 2018, pp. 3403--3407.

\bibitem{ClassificationQatab}
Bassam~Ali Al-Qatab and Mumtaz~Begum Mustafa,
\newblock ``Classification of dysarthric speech according to the severity of impairment: an analysis of acoustic features,''
\newblock {\em IEEE Access}, vol. 9, pp. 18183--18194, 2021.

\bibitem{PhonologicalRudzicz}
Frank Rudzicz,
\newblock ``Phonological features in discriminative classification of dysarthric speech,''
\newblock in {\em 2009 IEEE International Conference on Acoustics, Speech and Signal Processing}, 2009, pp. 4605--4608.

\bibitem{OROZCOARROYAVE2018207}
Juan~Rafael Orozco-Arroyave, Juan~Camilo Vásquez-Correa, Jesús~Francisco Vargas-Bonilla, R.~Arora, N.~Dehak, P.S. Nidadavolu, H.~Christensen, F.~Rudzicz, M.~Yancheva, H.~Chinaei, A.~Vann, N.~Vogler, T.~Bocklet, M.~Cernak, J.~Hannink, and Elmar Nöth,
\newblock ``Neurospeech: An open-source software for parkinson's speech analysis,''
\newblock {\em Digital Signal Processing}, vol. 77, pp. 207--221, 2018,
\newblock Digital Signal Processing \& SoftwareX - Joint Special Issue on Reproducible Research in Signal Processing.

\bibitem{TorgoSchu}
Guilherme Schu, Parvaneh Janbakhshi, and Ina Kodrasi,
\newblock ``On using the ua-speech and torgo databases to validate automatic dysarthric speech classification approaches,''
\newblock in {\em ICASSP 2023 - 2023 IEEE International Conference on Acoustics, Speech and Signal Processing (ICASSP)}, 2023, pp. 1--5.

\bibitem{xue21c_interspeech}
Wei Xue, Roeland van Hout, Fleur Boogmans, Mario Ganzeboom, Catia Cucchiarini, and Helmer Strik,
\newblock ``{Speech Intelligibility of Dysarthric Speech: Human Scores and Acoustic-Phonetic Features},''
\newblock in {\em Proc. Interspeech 2021}, 2021, pp. 2911--2915.

\bibitem{boersma2001praat}
Paul Boersma,
\newblock ``Praat, a system for doing phonetics by computer,''
\newblock {\em Glot. Int.}, vol. 5, no. 9, pp. 341--345, 2001.

\bibitem{Duffy_2020}
Joseph~R. Duffy,
\newblock {\em Motor speech disorders: Substrates, differential diagnosis, and management},
\newblock Elsevier, 2020.

\bibitem{kim2005dysarthria}
HyangHee Kim,
\newblock ``Dysarthria evaluation,''
\newblock {\em Communication Sciences \& Disorders}, pp. 23--28, 2005.

\bibitem{radford2022robust}
Alec Radford, Jong~Wook Kim, Tao Xu, Greg Brockman, Christine McLeavey, and Ilya Sutskever,
\newblock ``Robust speech recognition via large-scale weak supervision,''
\newblock {\em arXiv preprint arXiv:2212.04356}, 2022.

\bibitem{ip-detection}
Jeehyun Lee, Yerin Choi, Tae-Jin Song, and Myoung-Wan Koo,
\newblock ``Inappropriate pause detection in dysarthric speech using large-scale speech recognition,''
\newblock in {\em ICASSP 2024 - 2024 IEEE International Conference on Acoustics, Speech and Signal Processing (ICASSP)}, 2024, pp. 12486--12490.

\bibitem{bert-score}
Tianyi Zhang, Varsha Kishore, Felix Wu, Kilian~Q. Weinberger, and Yoav Artzi,
\newblock ``Bertscore: Evaluating text generation with {BERT},''
\newblock in {\em 8th International Conference on Learning Representations, {ICLR} 2020, Addis Ababa, Ethiopia, April 26-30, 2020}. 2020, OpenReview.net.

\bibitem{bert}
Jacob Devlin, Ming{-}Wei Chang, Kenton Lee, and Kristina Toutanova,
\newblock ``{BERT:} pre-training of deep bidirectional transformers for language understanding,''
\newblock in {\em Proceedings of the 2019 Conference of the North American Chapter of the Association for Computational Linguistics: Human Language Technologies, {NAACL-HLT} 2019, Minneapolis, MN, USA, June 2-7, 2019, Volume 1 (Long and Short Papers)}, Jill Burstein, Christy Doran, and Thamar Solorio, Eds. 2019, pp. 4171--4186, Association for Computational Linguistics.

\bibitem{bert-score-dysarthric}
Jimmy Tobin, Qisheng Li, Subhashini Venugopalan, Katie Seaver, Richard Cave, and Katrin Tomanek,
\newblock ``Assessing {ASR} model quality on disordered speech using bertscore,''
\newblock {\em CoRR}, vol. abs/2209.10591, 2022.

\bibitem{klue-bert}
Sungjoon Park, Jihyung Moon, Sungdong Kim, Won{-}Ik Cho, Jiyoon Han, Jangwon Park, Chisung Song, Junseong Kim, Youngsook Song, Tae~Hwan Oh, Joohong Lee, Juhyun Oh, Sungwon Lyu, Younghoon Jeong, Inkwon Lee, Sangwoo Seo, Dongjun Lee, Hyunwoo Kim, Myeonghwa Lee, Seongbo Jang, Seungwon Do, Sunkyoung Kim, Kyungtae Lim, Jongwon Lee, Kyumin Park, Jamin Shin, Seonghyun Kim, Eunjeong~Lucy Park, Alice Oh, Jung{-}Woo Ha, and Kyunghyun Cho,
\newblock ``{KLUE:} korean language understanding evaluation,''
\newblock in {\em Proceedings of the Neural Information Processing Systems Track on Datasets and Benchmarks 1, NeurIPS Datasets and Benchmarks 2021, December 2021, virtual}, Joaquin Vanschoren and Sai{-}Kit Yeung, Eds., 2021.

\bibitem{wpm}
Aihong Du, Chundan Lin, and Jingjing Wang,
\newblock ``Effect of speech rate for sentences on speech intelligibility,''
\newblock {\em 2014 IEEE International Conference on Communication Problem-Solving, ICCP 2014}, pp. 233--236, 03 2015.

\bibitem{agtabular}
Nick Erickson, Jonas Mueller, Alexander Shirkov, Hang Zhang, Pedro Larroy, Mu~Li, and Alexander Smola,
\newblock ``Autogluon-tabular: Robust and accurate automl for structured data,''
\newblock {\em arXiv preprint arXiv:2003.06505}, 2020.

\bibitem{adamw}
Ilya Loshchilov and Frank Hutter,
\newblock ``Decoupled weight decay regularization,''
\newblock in {\em International Conference on Learning Representations}, 2017.

\bibitem{yeo2022cross}
Eun~Jung Yeo, Kwanghee Choi, Sunhee Kim, and Minhwa Chung,
\newblock ``Cross-lingual dysarthria severity classification for english, korean, and tamil,''
\newblock in {\em 2022 Asia-Pacific Signal and Information Processing Association Annual Summit and Conference (APSIPA ASC)}. IEEE, 2022, pp. 566--574.

\bibitem{parselmouth}
Yannick Jadoul, Bill Thompson, and Bart de~Boer,
\newblock ``Introducing {P}arselmouth: A {P}ython interface to {P}raat,''
\newblock {\em Journal of Phonetics}, vol. 71, pp. 1--15, 2018.

\bibitem{baevski2020wav2vec}
Alexei Baevski, Yuhao Zhou, Abdelrahman Mohamed, and Michael Auli,
\newblock ``wav2vec 2.0: A framework for self-supervised learning of speech representations,''
\newblock {\em Advances in neural information processing systems}, vol. 33, pp. 12449--12460, 2020.

\bibitem{rathod23_interspeech}
Siddharth Rathod, Monil Charola, Akshat Vora, Yash Jogi, and Hemant~A. Patil,
\newblock ``{Whisper Features for Dysarthric Severity-Level Classification},''
\newblock in {\em Proc. INTERSPEECH 2023}, 2023, pp. 1523--1527.

\end{thebibliography}

\end{document}